\documentclass[twocolumn,superscriptaddress,floatfix,prl,longbibliography]{revtex4-1}
\usepackage{graphicx}
\usepackage{dcolumn}
\usepackage{amsmath}
\usepackage{amssymb}
\usepackage{amsbsy}
\usepackage{units}
\usepackage{color}
\usepackage{xcolor}
\usepackage{ulem}
\usepackage{textcomp,txfonts}

\begin{document}

\title{Fragile altermagnetism and orbital disorder in Mott insulator LaTiO$_3$}

\author{I. V. Maznichenko}
\email{igor.maznichenko@physik.uni-halle.de}
\affiliation{Department of Engineering and Computer Sciences, Hamburg University of Applied 
             Sciences, Berliner Tor 7, D-20099 Hamburg, Germany}
\affiliation{Institute of Physics, Martin Luther University Halle-Wittenberg,
             D-06099 Halle, Germany}

\author{A. Ernst}
\affiliation{Institute for Theoretical Physics, Johannes Kepler University, A-4040 Linz, Austria}
\affiliation{Max Planck Institute for Microstructure Physics, Weinberg 2, D-06120 Halle, Germany}

\author{D. Maryenko}
\affiliation{RIKEN Center for Emergent Matter Science (CEMS), Wako 351-0198, Japan}

\author{V. K. Dugaev}
\affiliation{Department of Physics and Medical Engineering, Rzesz\'ow University of Technology, 
 35-959 Rzesz\'ow, Poland}

\author{E.~Ya.~Sherman}
\affiliation{Department of Physical Chemistry and the EHU Quantum Center, University of the Basque Country UPV/EHU, Bilbao 48080, Spain}
\affiliation{Ikerbasque, Basque Foundation for Science, Bilbao 48013, Spain}

\author{P. Buczek}
\affiliation{Department of Engineering and Computer Sciences, Hamburg University of Applied 
             Sciences, Berliner Tor 7, D-20099 Hamburg, Germany}

\author{S. S. P. Parkin}
\affiliation{Max Planck Institute for Microstructure Physics, Weinberg 2, D-06120 Halle, Germany}

\author{S. Ostanin}
\affiliation{Institute of Physics, Martin Luther University Halle-Wittenberg,
             D-06099 Halle, Germany}

\begin{abstract}
 Based on {\it ab initio} calculations, we  {demonstrate} that a Mott insulator 
 LaTiO$_3$ (LTO), not {inspected} previously as an altermagnetic material, shows 
 the characteristic features of altermagnets, i.e.,  (i) fully compensated antiferromagnetism 
 and (ii) $\mathbf{k}$-dependent spin-split electron bands in the absence of spin-orbit coupling. 
 The altermagnetic ground state of LTO is protected by the crystal symmetry
 and specifically ordered $d$-orbitals of Ti ions with the orbital momentum $l=2.$
 The altermagnetism occurs when sites of Ti pair in the unit cell are occupied 
 by single electrons with $m=-1,s_{z}=+1/2$ and $m=+1,s_{z}=-1/2$ per site,
 with $m$ and $s_{z}-$ being the $z-$ component of the orbital momentum and spin, respectively.
 By further simulating orbital disorder within the Green's function method, 
 we disclose its damaging character on the spin splitting and the resulting altermagnetism.  
 When the single-electron spin-polarized state at each Ti site is contributed almost 
 equally by two or three $t_{2g}$ orbitals, LTO becomes antiferromagnetic.  
 The effect of the spin-orbit coupling, which can cause orbital disorder and 
 suppress altermagnetism, is discussed.  
\end{abstract}

\maketitle

\section{Introduction}
\label{sec:introduction}

  Altermagnets (AM) are a subset of the large class of collinear antiferromagnetic 
 materials which display no net magnetization yet which have $k$-dependent spin-split 
 electronic bands. This is in contrast with conventional collinear antiferromagnets 
 for which the up and down spin polarized bands are indistinguishable throughout $k$ 
 space.\cite{Smejkal-PRX12-2022, Mazin-Physics, Yuan-AdvMater-2023}
%% Lin-Ding Yuan and Alex Zunger, Degeneracy Removal of Spin Bands ...,
%% Adv. Mater. 2023, 35, 2211966  ; DOI: 10.1002/adma.202211966. 
%% L. Šmejkal, J. Sinova, and T. Jungwirth, Beyond conventional ferromagnetism and 
%% antiferromagnetism: A phase with nonrelativistic spin and crystal rotation symmetry,
%% Phys. Rev. X 12, 031042 (2022).
%% I. Mazin (The PRX Editors), Editorial: Altermagnetism — a new punch line of fundamental 
%% magnetism, Phys. Rev. X 12, 040002 (2022).
 Using {\it ab initio} electronic band structure calculations, within the density functional theory
 (DFT) and a nonrelativistic spin group formalism \cite{NonRel-SpinGroup},   
%% D. B. Litvin, Spin Point Groups, Acta Crystallogr. Sect. A33, 279 (1977).
  more than a dozen candidate altermagnets have been proposed, including insulating 
 fluorides (MnF$_2$), oxides (Fe$_2$O$_3$) and perovskites (LaMnO$_3$),  chalcogenide 
 semiconductors (MnTe), semimetals (CoNb$_3$S$_6$), and various metals (RuO$_2$, CrSb, 
 Mn$_5$Si$_3$).\cite{Smejkal-PRX12-040501, Spaldin2024}
%% L. Šmejkal, J. Sinova, and T. Jungwirth, Emerging Research Landscape of Altermagnetism,
%% Phys. Rev. X 12, 040501 (2022).    
 The crystal and electronic structures of these proposed AMs, as well as 
 their chemical compositions and properties are diverse. Yet, all AMs exhibit
 time-reversal ($\mathcal{T}$) symmetry breaking which leads to a spin-split electronic
 structure in the absence of spin-orbit coupling.\cite{Yuan-PRMater-2021}
%%Lin-Ding Yuan, Zhi Wang, Jun-Wei Luo, and Alex Zunger, Prediction of low-Z collinear and ...
%%Phys. Rev. Materials 5, 014409 (2021). 
 This leads to a number of spin-dependent phenomena that make AMs of potential interest for 
 spintronics applications.\cite{Smejkal-PRX12-040501, Naka-PRB-2021}
%% Makoto Naka, Yukitoshi Motome, and Hitoshi Seo, Perovskite as a spin current generator,
%% Phys. Rev. B 103, 125114 (2021). 
 For the case of RuO$_2$, 
 the {band structure predictions are supported by observations of an anomalous Hall effect 
 \cite{Z.Feng_NatEl.2022}, 
%% Zexin Feng et al. An Anomalous Hall Effect in Altermagnetic Ruthenium Dioxide,
%%  Nat. Electron. 5, 735 (2022).
 spin current and torque \cite{A.Bose_NatEl.2022, H.Bai_PRL_128_197202, S.Karube_PRL_129_137201}
 as well as the observation of magnetic circular 
 dichroism  in  angle-resolved photoemission spectroscopy (ARPES).\cite{O.Fedchenko_SciAdv_10_eadj4883}}
%% Arnab Bose et al. Tilted Spin Current Generated by an Antiferromagnet, 
%% Nat. Electron. 5, 263 (2022).
%% Hua Bai et al. Phys. Rev. Lett. 128, 197202 (2022).
%% Shutaro Karube et al. Phys. Rev. Lett. 129, 137201 (2022).
 AMs can show relatively high magnetic ordering temperatures {$T_{\rm AM}$} and 
 large spin-splitting of the electronic bands which has stimulated experimental efforts to provide direct  
 evidence of altermagnetism from ARPES measurements. Indeed,
 spin-splitting of the electronic bands in $\alpha$-MnTe ({$T_{\rm AM}$}=267~K)
 have been recently reported to be as much as $\sim$370~meV.\cite{S.Lee_PRL_132_036702, J.Krempasky_Nat_626_517}
%% Suyoung Lee et al. Broken Kramers Degeneracy in Altermagnetic MnTe,
%% PHYSICAL REVIEW LETTERS 132, 036702 (2024).
%% J. Krempaský et al. Nature 626, 517–522 (2024).
%% Olena Fedchenko et al. Observation of time-­reversal symmetry breaking in the band 
%% structure of altermagnetic RuO2, Sci. Adv. 10, eadj4883 (2024).

  Here we consider the Mott insulator LaTiO$_3$  (LTO) that has not previously been 
 calculated as an altermagnet, although the material meets needed 
 requirements.\cite{Smejkal-PRX12-040501, Yuan-AdvMater-2023} 
 (i) LTO is orthorhombic with a unit cell that contains two antiferromagnetically 
 $G$-ordered sublattices of Ti$^{3+}$.
 (ii) There is no inversion center between the Ti sites because of tilted TiO$_6$ octahedra.
 (iii) The opposite-spin sublattices are connected by translation combined with rotation. 
 In LTO, the semiconducting band gap separates the occupied Ti $t_{2g}^{1} e_g^{0}$ subband 
 from the unoccupied conduction band edge, while the $f$ states of La appear at much 
 higher energies. The key structural factor of $Pbnm$-LTO, which determines its band gap 
 and zero magnetization, is optimally rotated TiO$_6$.\cite{Maznichenko-arxiv2024}
  The symmetry of LTO does not differ from that of other {\it AB}O$_3$ potential 
 altermagnets, such as LaVO$_3$ or LaMnO$_3$. 
 However, the occupied Ti-based $d$ subband in LTO contains one electron only, i.e., 
 the lowest possible value. This single orbital filling enlarges the  
 spin-split effects and their visibility in the bandstructure, producing a
 clear picture of the altermagnetism appearance.

  In this work, we show from first principles calculations that orthorhombic LTO exhibits 
 an altermagnetic band structure, focusing on the 
 Ti $d$-orbital filling. {In cubic LTO, there are three $t_{2g}$ orbitals which can be 
 occupied by one electron.} Quantum effects such as a Jahn-Teller distortion and spin-orbit 
 coupling can remove the $t_{2g}$ degeneracy. 
 Keimer {\it et al.} \cite{B.Keimer_PRL_85_3946} 
%% B. Keimer et al., PRL 85 3946 2000.
 reported that the orbital momentum of the $t_{2g}$ level is quenched, and suggested a picture  
 of strongly fluctuating $t_{2g}$ orbitals. 
 The observed anomalies were explained 
 by assuming the formation of an orbital liquid state.\cite{G.Khaliullin_PRL_85_3950} 
%% G. Khaliullin and S. Maekawa, Orbital Liquid in Three-Dimensional Mott Insulator: LaTiO 3, 
%% PRL 85 3950 2000.
 Keeping in mind the scenario of orbital fluctuations, we {further} present here   
 simulations of orbital disorder, using a Green's function method.
 First of all, for the ordered and specifically filled Ti $t_{2g}$ orbitals,
 our DFT calculations show a strongly split spin bands.  
% \ES{\sout{Theoretically, the LTO altermagnetism seems reliable.}}
 Then, to simulate orbital disorder, the charge of one electron was distributed 
 among the two or three $t_{2g}$ orbitals on each Ti site 
 to show the important effects of orbital mixing on the spin-splitting of the electron bands. We conclude that when 
 at least two $d$ orbitals of Ti become almost equally filled, altermagnetic LTO transforms 
 into an antiferromagnet.

%%%%%%%%%%%%%%%%%%%%%%%%%%%%%%%%%%%%%%%%%%%%%%%%%%%%%%%%%%%%%%%%%%%%%%%%%%%%%%%%%%

\section{Details of calculations}
\label{sec:Details}

  The electronic and magnetic properties of LTO were computed from first 
 principles using the {Korringa-Kohn-Rostoker (KKR)} Green's function method \cite{Luders, Geilhufe}
%% M. Lüders, A. Ernst, M. Däne, Z. Szotek, A. Svane, D. Ködderitzsch, W. Hergert, 
%% B. L. Györffy, and W. M. Temmerman, Phys.Rev. B 71, 205109 (2005).
%% M. Geilhufe, S. Achilles, M. A. Köbis, M. Arnold, I. Mertig,
%% M. Hergert, and A. Ernst, J. Phys.: Condens. Matter 27, 435202 (2015).
 based on multiple scattering theory within the DFT and generalized gradient 
 approximation  \cite{GGA-PBE}  to the exchange-correlation potential.
%% J. P. Perdew, K. Burke, and M. Ernzerhof, Phys. Rev. Lett. 77, 3865 (1996).
 The full charge-density treatment used here takes into account a nonsphericity 
 of both the crystal potential and charge density. 
 The maximal angular momentum used was {$l_{\rm max}$}=3 and the integrals over the Brillouin 
 zone were performed using a {$16\times 16\times 16$} $k$ mesh.

 For the AM configurations of LTO, the disordered filling of the Ti ($l=2, m, s_{z}$) 
 states was simulated using the coherent potential approximation (CPA)
  \cite{Gyorffy_PRB-5-2382, Oguchi_MetPhys-13-145, Gyorffy_JPhysF-15-1337}. 
%% B. L. Gyorffy, Phys. Rev. B 5, 2382 (1972).
%% [50] T. Oguchi, K. Terakura, and N. Hamada, J. Phys. F: Met. Phys. 13, 145 (1983).
%% [51] B. L. Gyorffy, A. J. Pindor, J. Staunton, G. M. Stocks, and H. Winter, 
%% J. Phys. F: Met. Phys. 15, 1337 (1985).
%% The use of CPA allows us to simulate also the presence of vacancies on the La and oxygen 
%% sites, which may appear during the sample preparation. 
%% The defect structures are denoted as La$_{1−\delta}$TiO$_{3}$ and LaTiO$_{3−\delta}$,
%% where $\delta$ varies between zero and 0.05 at.~\% . 
 The method adequately mimics homogeneously distributed disorder 
  without a large supercell.

\begin{figure}
   \centering
   \includegraphics[width = 1\columnwidth]{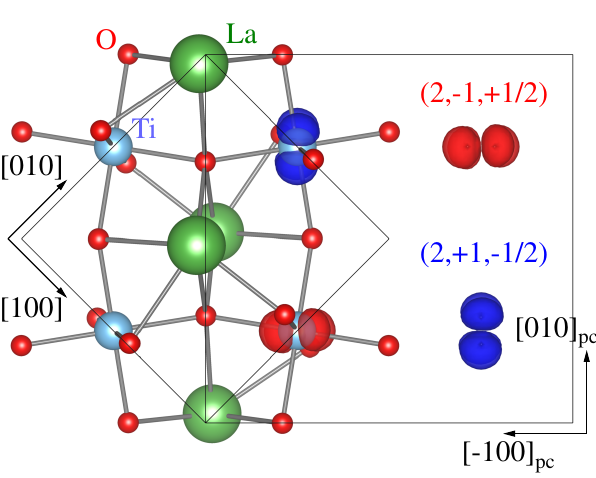}
   \caption{Top view of the magnetic unit cell, $\sqrt{2}a \times \sqrt{2}b$, of orthorhombic 
    LaTiO$_3$.  The ordered and filled orbitals of a pair of Ti atoms
  ($l=2,m=-1,s_{z}=+1/2$) and ($l=2,m=+1,s_{z}=-1/2$) are shown, respectively, by red and blue color. 
  The directions of the pseudocubic cell are labeled as 'pc'.}
  \label{cell}
\end{figure}

 To reproduce the LTO properties within the DFT+$U$ parametrization \cite{DFT+U}, 
  correlation parameters {$U_{\rm eff}$} of 2.3 eV and 6 eV, were applied to the Ti 3$d$ 
 states and $f$ states of La, respectively.  
 The $D_{2h}$ magnetic unit cell of LTO, with four formula units (f. u.) in the $a-b$ plane, 
 is  shown in Figure~\ref{cell}. The cell allows the modeling of the TiO$_6$
octahedral tilting, 
 antiferromagnetically ordered Ti and the altermagnetic band structure. 

\section{Results and discussion}

%%%%%%%%%%%%%%%%%%%%%%%%%%%%% DOS %%%%%%%%%%%%%%%%%%%%%%%%%%%%%%%%%%%%%%%%

\begin{figure}
   \centering
   \includegraphics[width = 1.05\columnwidth]{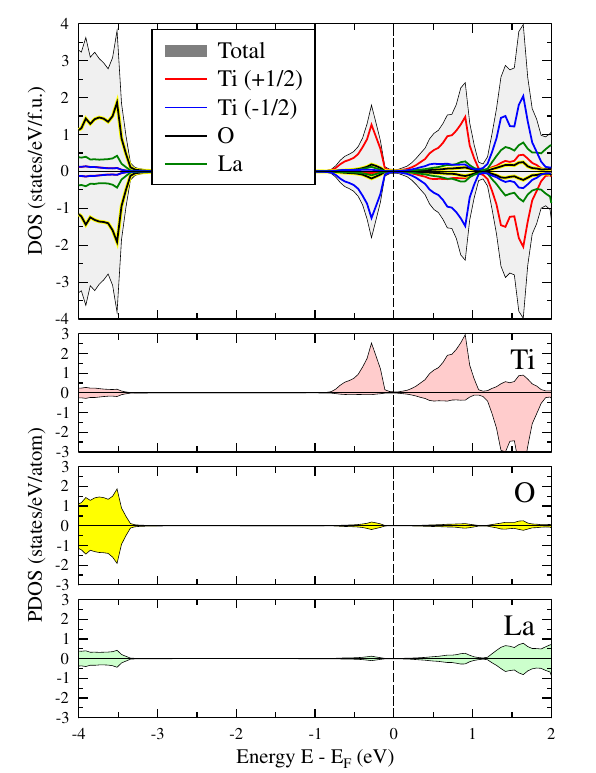}
%   \caption{Total and spin-polarized DOS of orthorombic LaTiO$_3$ (LaTi$\uparrow_0.5$Ti$\downarrow_{0.5}$O$_3$ ) per formula unit.}
   \caption{ The total spin-resolved density of states (DOS) and its site-projected components (PDOS) of orthorhombic LaTiO$_3$ per formula unit and 
   per atom, respectively.}
  \label{dos}
\end{figure}
 The total spin-resolved density of states (DOS) and its site-projected components (PDOS) 
 are shown in Figure~\ref{dos}.  A band gap of 
 0.22 eV separates the occupied Ti $d$ subband from the unoccupied conduction bands,
 the bottom edge ($E_c$) of which is also formed by Ti orbitals. The $f$ states of La appear well
 above $E_c$ by $ + 1.3$~eV. Since each Ti$^{3+}$ has an antiferromagnetically (AFM) ordered partner, 
 we show in Figure~\ref{dos} only one Ti.
 According to our calculations, each Ti$^{3+}$ has a magnetic moment with a value of {$\sim0.8 {\mu}_{\rm B}$}.

%%%%%%%%%%%%%%%%%%%%%%%%%%%%% Energetics %%%%%%%%%%%%%%%%%%%%%%%%%%%%%%%%%%%%%

  First, we calculated all the ordered electronic configurations where  
  only one $t_{2g}$ orbital of each Ti is completely filled.
 Since the ($l=2, m, s_{z}$) nomenclature is very suitable for the 3$d$ Ti states 
 of orthorhombic LTO, this orbital nomenclature is used below.
 For comparison, the Ti states ($2, m=-1, s_{z}$) and ($2, m=+1, s_{z}$) can be associated 
 with those denoted in cubic perovskites as $d_{xz}$  and $d_{yz}$, respectively.
 The nomenclature ($2, -2, s_{z}$) corresponds to $d_{xy}$.
 The total energies computed for the three AM, FM  and three differently
 ordered AFM configurations of LTO are shown in Table~\ref{tab1}.
 By inspecting these energies we find that the $G$-type AM structure has the lowest energy.
 The AM-$A$ and AM-$C$ structures are not favorable energetically, as compared to
 that of AM-$G$. 
% In LTO, its altermagnetism is protected not only by broken symmetry. \ES{by what else ?}
 Additionally, to keep the 4-f.u.-cell LTO rigorously altermagnetic, the filled 
 orbital nomenclature {($l,m,s_{z}$)} of two Ti$^{3+}$ should be (2,-1,+1/2), while the one-orbital 
 filling of  two other Ti is (2,+1,-1/2). 
 Therefore, the ordered filling of the Ti $t_{2g}$ orbitals with magnetic quantum numbers 
 $m=\pm1$ plays a key role for the emergence of AM in LTO.  
    
\begin{table}
\caption{Total energies of  LTO per formula unit, calculated for the FM, three AFM
 ($A$-, $C$- and $G$-type), three AM configurations and the orbital-ordered (OO) FM configuration. Each energy value is shown 
 with respect to the AM-$G$ ground state. The one-orbital filled states ($l, m, s_{z}$) 
 are given for each Ti pair.}
% and each magnetic configuration.}
\begin{center}
\begin{tabular}{|c|c|c|}
\hline
Configuration & Filled states of Ti pair & Total energy (meV/f.u.) \\
\hline
AFM $A$   & (2,-2,+1/2),(2,-2,-1/2)  & 92.89 \\
\hline
AFM $C$   & (2,-2,+1/2),(2,-2,-1/2)  & 114.60 \\
\hline
AFM $G$   & (2,-2,+1/2),(2,-2,-1/2)  & 115.41 \\
\hline
FM      &  (2,-2,+1/2),(2,-2,+1/2) & 115.21 \\
\hline
AM $A$    & (2,-1,+1/2),(2,+1,-1/2)  & 59.64 \\
\hline
AM $C$    & (2,-1,+1/2),(2,+1,-1/2)  & 115.11 \\
\hline
AM $G$    & (2,-1,+1/2),(2,+1,-1/2)  & 0.00 \\
\hline
FM(OO)  & (2,-1,+1/2),(2,+1,+1/2)  & 15.30 \\
\hline
\end{tabular}
\end{center}
\label{tab1}
\end{table}

%%%%%%%%%%%%%%%%%%%%%%%%%%%%%% Altermagnetic band structure %%%%%%%%%%%%%%%%%%%%%%

\begin{figure}
   \centering
   \includegraphics[width = 1.\columnwidth]{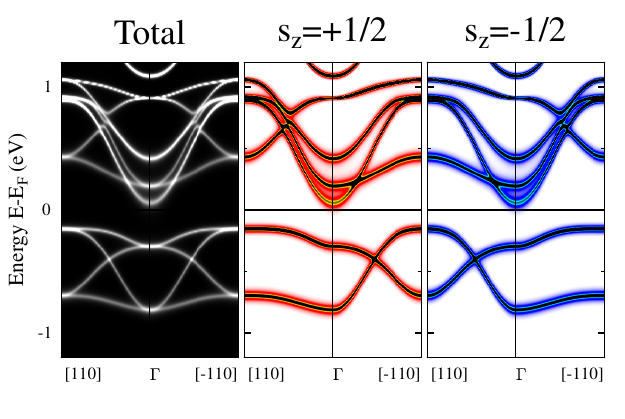}
   \caption{Total and spin-resolved Bloch spectral functions of altermagnetic LTO,
     which are plotted along the [110] and [-110] $k$-directions near $E_{F}=0$
     between -1.2~eV$< E - E_{F} <$+1.2~eV.}
  \label{bandstructure}
\end{figure}
  In Figure~\ref{bandstructure}, the nonrelativistic total and spin-resolved Bloch spectral 
 functions of energetically preferable AM-$G$ configuration of LTO 
 are plotted along the [110] and [-110] directions of the Brillouin zone (BZ). 
 These correspond to [100]$_{pc}$ and [010]$_{pc}$ for the pseudocubic cell.
 We show in Figure~\ref{bandstructure} the branches $E(\mathbf{k})$, which appear in 
 the energy window of 2.4~eV around the Fermi level ($E_{F} =0$). 
 The low conduction bands of LTO, seen in Figure~\ref{bandstructure} above $E_{F}$ 
 are related solely to the unoccupied Ti 3$d$ states.
 The bands seen in Figure~\ref{bandstructure} below $E_{F}$
 represent the filled states ($l=2, m=-1$) for $s_{z}=+1/2$ and  ($l=2, m=+1$) for $s_{z}=-1/2.$  
 The total Bloch spectral function, plotted in the left panel of 
 Fig.~\ref{bandstructure}, shows that each $E(\mathbf{k})$ has spin 
 degeneracy and gives an idea how the antiferromagnetic band structure of LaTiO3 looks like.
%%shows, in fact, an antiferromagnetic band structure, 
%%where each $E(\mathbf{k})$ has spin degeneracy. 
 For comparison, the two spin-resolved $E(\mathbf{k})$, which 
 are plotted in the middle and right panels of Figure~\ref{bandstructure}, represent 
 the spin-spit AM band structure. Within this approach altermagnetism in LTO 
 is strongly energetically favored.

%%%%%%%%%%%%%%%%%%%%%%%%%%%%%%%%%%% (001) cross sections %%%%%%%%%%%%%%%%%%%%%%%%%%%%%%%

\begin{figure}
  \centering
  \includegraphics[width = 1\columnwidth]{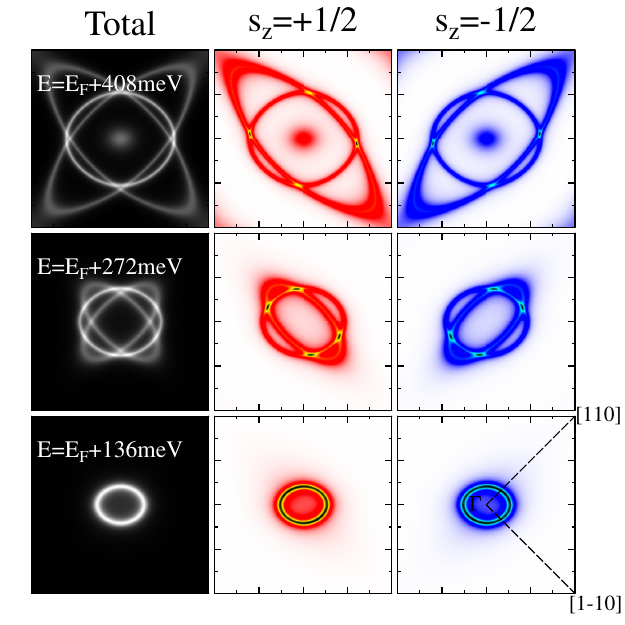}
  \caption{Total and spin-resolved conduction-band cross sections of
  altermagnetic LaTiO$_3$, which were calculated within the BZ plane ${k}_z = 0$ at
  $E = E_{F} +0.14$ eV, $E = E_{F} +0.27$ eV and $E = E_{F} +0.41$ eV.}
  \label{CB-cross-sections}
\end{figure}
 % Since LTO is a narrow-gap insulator, its surface could become conducting for example, 
 % due to oxygen vacancies that could arise intrinsically or which could be introduced by 
 % bias fields (external or from band bending induced by another compound) or by chemical doping.
 A narrow gap Mott insulator LTO can become conducting \cite{Maryenko-APLMat2023}
 without a strong modification of its crystal structure. 
 The conductivity origins can include decreased tilting angles of TiO$_6$ octahedra \cite{Maznichenko-arxiv2024}, 
 a small concentration  of oxygen vacancies providing electron doping without distorting the structure, and 
 an applied external bias. Here, for altermagnetic LTO, we mimic its 
 hypothetical metallization by assuming that an external bias is applied.
 Cross sections of $E(\mathbf{k})$ within the BZ plane {$k_{z} = 0$} are 
 plotted in Figure~\ref{CB-cross-sections}. At $E = E_{F} +$140~meV, 
 the Fermi surface (FS) would consist of a few relatively small %and isotropic 
 electron sheets closed around the BZ centre $\Gamma$, as shown in the lower panels of 
 Figure~\ref{CB-cross-sections}. Then, with increasing bias voltage, all FS sheets 
 enlarge and, simultaneously, extend along [1,1] and [-1,1] within 
 the {$k_{z} = 0$} plane. For energies which exceed $E_{F}$ by more than 0.3~eV,
 we find that the total FS has a fourfold symmetry, whereas each spin contributes
 to twofold sheets perpendicularly extended to each other.   
 The spin-resolved FS {contour lines} of altermagnetic LTO, which is shown 
 in the upper panels 
 of Figure~\ref{CB-cross-sections}, suggests that rather strong in-plane anisotropy can 
 appear in the spin-dependent conductivity at 0.3--0.4 eV above $E_{F}$.  
 This is due to altermagnetism and appears when the bias {or doping}
 probes the conduction bands. {It should be noted that the FS {contour lines} 
 shown in Figure~\ref{CB-cross-sections} is related to the $Pbnm$-LTO~(001) plane.}

\begin{figure}
  \centering
  \includegraphics[width = 1.\columnwidth]{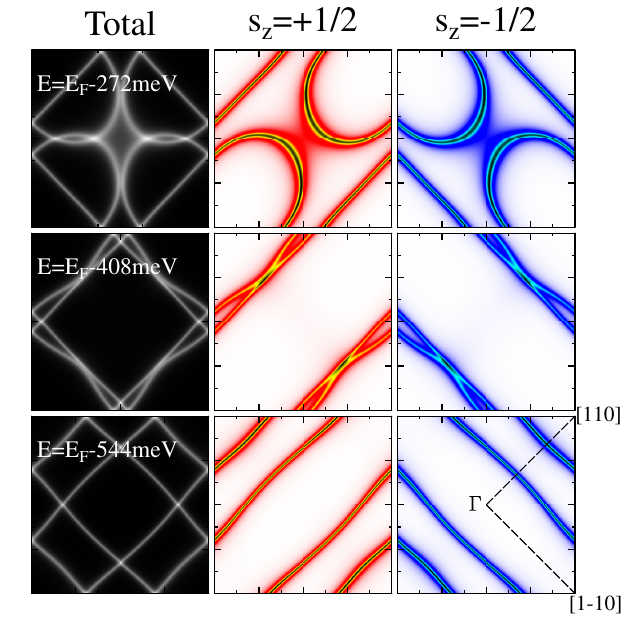}
  \caption{Total and spin-resolved valence-band cross sections of
  altermagnetic LaTiO$_3$, which were calculated within the BZ plane $k_z = 0$ at 
  $E = E_{F} -0.27$ eV, $E = E_{F} -0.41$ eV, and $E = E_{F} -0.54$ eV.} 
  \label{VB-cross-sections}
\end{figure}
 The use of an appropriate {shift of the Fermi level} allows one to also consider the occupied Ti $t_{2g}$ subband 
 of altermagnetic LTO. Three valence-band cross sections, calculated within the 
 BZ plane {$k_{z}=0$} at $E = E_{F} -0.27$ eV, $E = E_{F} -0.41$ eV, and $E = E_{F} -0.54$ eV, 
 are plotted in Figure~\ref{VB-cross-sections}.  
 For each cut of $E(\mathbf{k})$, the total FS of fourfold symmetry includes spin-dependent 
 contributions, each of which appears as open sheets of twofold symmetry. Since the 
 spin-up and spin-down sheets are elongated perpendicularly to each other, as 
 Figure~\ref{VB-cross-sections} shows, we expect that the in-plane spin transport in 
 altermagnetic LTO might be highly anisotropic. 
 % \ES{\sout{under negative bias voltage.}} 
 Again, these predictions relate to the LTO~(001) {plane} only. 
 %\ES{\sout{Moreover, 
% LTO is expected to be a single-domain material that seems questionable.} Better in Discussion.} 

%%%%%%%%%%%%%%%%%%%%%%%%%%%%% Orbital disorder  %%%%%%%%%%%%%%%%%%%%%%%%%%%%%%%%

%\begin{figure}
%   \centering
%   \includegraphics[width = 1.0\columnwidth]{bands5xzyzxy05allUP.pdf}
%   \caption{Spin-resolved band structures of LTO with 
%  disorderedly filled Ti orbitals. The (2,-2,$\pm1/2$) filling of 0.1~$e$ and 0.5~$e$ 
%  is plotted in the upper and low panels, respectively.}
%  \label{CPA-bandstructure}
%\end{figure}
\begin{figure}
   \centering
   \includegraphics[width = 1.0\columnwidth]{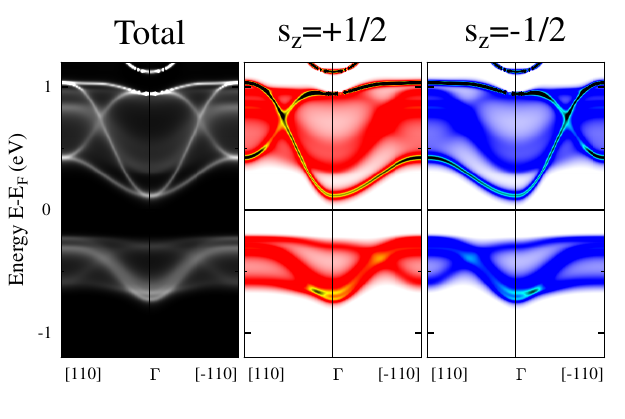}
   \caption{Spin-resolved band structure of LTO, calculated with the two equally filled  
   and disordered $d$ orbitals on each Ti site. Here, the initially AM (2,$\pm1$,$\pm1/2$) 
   orbitals and initially AFM (2,-2,$\pm1/2$) orbitals of each Ti were mixed and filled 
   by 0.5~$e$.}
  \label{CPA5050-bandstructure}
\end{figure}

Next, by simulating disorder for the Ti $t_{2g}$ orbitals with
the CPA method, we show the fragility of this altermagnetism due to
multiorbital effects.  { This effects occur in the case of completely
  filled (2,-1,+1/2) and (2,+1,-1/2) orbitals for each Ti pair.} Using
the KKR-CPA technique, several electronic configurations with
underfilled (2,-1,+1/2) and (2,+1,-1/2) were simulated. The missing
charges were kept to take the (2,-2,+1/2) and (2,-2,-1/2)
orbitals. The charge transfer, $\Delta q$, to the (2,-2,$\pm1/2$)
orbitals varied from zero to 0.5~electron ($e$) on each Ti.  The
corresponding effect of the two equally filled Ti orbitals on the band
structure of LTO is shown in Figure~\ref{CPA5050-bandstructure}.  The
well mixed electronic configuration calculated with $\Delta q =0.5 e$
illustrates explicitly the trends of the elmination of altermagnetism by
  multiorbital filling.  One can expect that the main changes in
$E(\mathbf{k})$ should take place below $E_F$, since the charge was
redistributed between the filled Ti orbitals only. Indeed, with
increasing $\Delta q$, initially large spin splitting weakens.  As
Figure~\ref{CPA5050-bandstructure} shows, when $\Delta q$ approaches
0.5$e$ the band structure of LTO below $E_{F}$ {is more conventionally
  antiferromagnetic than altermagnetic.}

\begin{figure}
   \centering
   \includegraphics[width = 1.0\columnwidth]{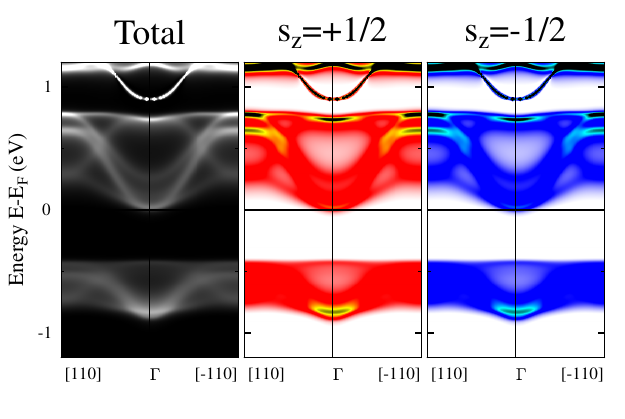}
   \caption{Spin-resolved band structure of the AFM $G$-type LTO, calculated with
  the three equally filled by 1/3 and disordered $t_{2g}$ orbitals of each Ti.}
  \label{CPA-bandstructure}
\end{figure}

 Finally, we simulated the case of completely disordered $t_{2g}$ orbitals {by taking three orbitals into consideration.} 
 The three orbitals of each Ti site, namely, (2,-2,$\pm1/2$), (2, 1,$\pm1/2$), and 
 (2,-1,$\pm1/2$) were filled equally by 1/3. The corresponding spin-polarized 
 band structure is plotted in Figure~\ref{CPA-bandstructure} which shows that the material now becomes a conventional AFM.

 The effects of orbital fluctuations, which can be triggered in LTO by spin-orbit 
 coupling (SOC) were studied, e.g., in \cite{B.Keimer_PRL_85_3946, G.Khaliullin_PRL_85_3950} 
 prior to the nonrelativistic concept of altermagnetism in the absence of any SOC. 
 Note that in our simulations, the SOC is also not considered. With inclusion of SOC, this relativistic effect renders spins and orbitals differently mixed for each 
 electronic band state $E({\bf k}),$ thus permitting orbital fluctuations and 
 making the SOC the main damaging factor of altermagnetism.

\section{Summary and outlook}

 We presented {\it ab initio} calculations of altermagnetic LaTiO$_3$ performed with the
 Green's function method and coherent potential approximation.
 First, we found that the energetically preferable 
 altermagnetic configuration is protected by the unit cell symmetry and, most importantly, by 
 specifically filled Ti $t_{2g}$ orbitals. For the Ti pairs in each unit cell, 
 the single-electron electron states ($l=2,m=-1,s_{z}=+1/2$) and ($l=2,m=+1,s_{z}=-1/2$) per site should be 
 filled. Second, we observed that orbital disorder 
 damages the spin splitting and, therefore, destroys the altermagnetism. 
 This happens when at least two orbitals, for instance, ($l=2,m=\pm1$) and ($l=2,m=-2$), 
 contribute equally to single-electron states at Ti sites. 
 In fact, superpositions of partially occupied Ti $t_{2g}$ orbitals 
 producing their total charge equal to one, support the antiferromagnetic configurations.  

 One of the key factors which can intrinsically provide orbital disorder in LaTiO$_3$ is the spin-orbit 
 coupling, not included in the basic nonrelativistic concept of altermagnetism. 
 We note that this relativistic effect can induce spin-split bands due to noncollinear magnetism,
 as has recently been observed for a noncoplanar antiferromagnet MnTe$_{2}$ \cite{Yu-Peng.Zhu_Natur_626_523},
 making the studies of the relationship between altermagnetism and noncollinear magnetism and
 between altermagnetism and spin-orbit coupling very interesting. \citep{hajlaoui2024temperature}
%%Zhu, Y.-P. et al. Nature 626, 523–528 (2024). 
  
 The altermagnetic band structure of LaTiO$_3$ that we have calculated can shed light on various 
 features of bulk LaTiO$_3$ and LaTiO$_3-$related heterostructures. First, its potentially anisotropic 
 charge and spin transport properties can be modeled based on the calculated bandstructure.   
 Second, the key element of the unit cell structure causing the Mott insulator behavior is
 the optimal tilting of the TiO$_{6}$ octahedra. This optimal structure realization is lifted at the {LaTiO$_3$}/KTaO$_3$ 
 interfaces leading to the recently reported \cite{Maryenko-APLMat2023} emergence of interface symmetry-dependent 
 two-dimensional conductivity and superconductivity. In this context, it would be worthwhile to investigate the role of altermagnetism 
 in formation of the optimal tilting and, in turn, the interplay between the bulk altermagnetism and 
the formation of the interface-based electron states.  
An important finding is that small
structural modifications of LTO can change the material from
an altermagnet to a conventional magnetics allowing
for the possibility of controllable switching of the spin-polarization of conduction electrons.

%\vspace{2cm}
%\section{Acknowledgments}
\begin{acknowledgments}
Authors thank Igor Mazin for useful discussion.
P.A.B. and A.E. acknowledge the funding by the Fonds zur Förderung der Wissenschaftlichen Forschung (FWF) under Grant No. I 5384. 
The work of E.S. is financially supported through the Grant PGC2018-101355-B-I00 funded by
MCIN/AEI/10.13039/501100011033 and by ERDF “A way
of making Europe”, and by the Basque Government through
Grant No. IT986-16.
\end{acknowledgments}

\bibliography{Altermagnetic-LTO}

\end{document}